\documentclass[%
reprint,
amsmath,amssymb,
aps,
onecolumn,
]{revtex4-2}

\usepackage{graphicx}% Include figure files
\usepackage{bm, slashed}% bold math
\usepackage{tikz} % tikz package
\usepackage{hyperref}% add hypertext capabilities
\usepackage{comment}
\usepackage{mathdots,bbold}
\usepackage{natbib}
\usepackage{soul}

\setcitestyle{square,numbers}

\begin{document}
	
	%\preprint{APS/123-QED}
	
	\title{Reformulation of %abelian 
		gauge theories in terms of gauge invariant fields}
	
	\author{Pierpaolo Fontana}
	\email{pfontana@sissa.it}
	\affiliation{%
		SISSA and INFN, Sezione di Trieste, Via Bonomea 265, I-34136 Trieste, Italy
	}%
	
	\author{Joao C. Pinto Barros}
	\email{jpintobarros@itp.unibe.ch}
	\affiliation{
		Albert Einstein Center for Fundamental Physics, Institute for Theoretical Physics, University of Bern, Sidlerstrasse 5, CH-3012 Bern, Switzerland
	}%
	
	\author{Andrea Trombettoni}
	\email{andreatr@sissa.it}
	\affiliation{%
		Department of Physics, University of Trieste, Strada Costiera 11, I-34151
		Trieste, Italy
	}%
	\affiliation{SISSA and INFN, Sezione di Trieste, Via Bonomea 265,
		I-34136 Trieste, Italy}
	
	\date{\today}
	
	\begin{abstract}
		We present a reformulation of %abelian 
		gauge theories in terms of gauge invariant fields. 
		Focusing on abelian theories, we show that the gauge and matter covariant fields can be recombined to introduce new gauge invariant degrees of freedom.
		Starting from the $(1+1)$ dimensional case on the lattice, with both periodic and open boundary conditions, we then generalize to higher dimensions and to 
		the continuum limit.
		To show explicit and physically relevant examples of the reformulation, we apply it to the Hamiltonian of a single particle in a (static) magnetic field,  to pure abelian lattice gauge theories, to the Lagrangian of quantum electrodynamics in $(3+1)$ dimensions and to the Hamiltonian of the $2d$ and $3d$ Hofstadter model. In the latter, we show 
		that the particular construction used to eliminate the the gauge covariant fields enters the definition of the magnetic Brillouin zone. Finally, we briefly comment on relevance of the presented reformulation to the study of interacting gauge theories.
	\end{abstract}
	
	\maketitle
	
	%\tableofcontents
	
	\section{\label{intro}Introduction}
	Due to their broad applicability, gauge theories have a key importance in physics that can be hardly overstated \cite{Peskin,Scwhartz,Maggiore}. In the field of particle physics they are at the basis of the Standard Model, a non-abelian gauge theory with gauge group $U(1)\times SU(2)\times SU(3)$, where the first two groups refer to the electroweak sector and the last to quantum chromodynamics (QCD). They also play an important role in
	condensed matter physics, where gauge fields may emerge from the effective description of strongly correlated phenomena at low energies, like quantum Hall systems and quantum spin liquids \cite{Wen}.
	
	The analysis of gauge theories in the strong coupling regime is an arduous problem where perturbative approaches typically fail. One way to deal with this problem is to work under the framework of lattice gauge theories (LGT)
	\cite{Wilson,Kogut-Susskind,Rothe,gattringer2009}. The lattice formulation at finite volume provides natural infrared and ultraviolet cut-offs that regularize the theory. Moreover, within this formulation, numerical approaches to the problem are possible using Monte Carlo methods \cite{Rothe,gattringer2009}	and crucial results have been obtained, e.g. for lattice QCD. Despite this success, there are various questions which remain intractable within the importance sampling Monte Carlo approach due to the well known sign or complex action problems. Examples include the study of high baryon density QCD or out of equilibrium real time evolution \cite{troyer_wiese2005,fukushima2010}.
		
	Recent developments in the field of the quantum simulation of many-body physics have brought attention to the Hamiltonian formulation of LGT \cite{Kogut-Susskind}. In principle, by engineering suitable local Hamiltonians, quantum simulators can be used to investigate problems that remain unsolved by classical computers. This, however, is still a complicated task: gauge theories entail a redundant description of nature where superfluous degrees of freedom are included in the model. This is manifested through the existence of local symmetries ensuring that the non-physical degrees of freedom decouple from the physical ones. The engineering of these symmetries is one of the great challenges of present day quantum simulators \cite{Wiese,Dalmonte-Montangero,Cirac-Zoller,Zohar-Cirac-Reznik,banuls2020,barros2020gauge}.
	
	The methodologies typically used to study gauge theories rely on the study of Lagrangians (or Hamiltonians) whose fundamental objects, i.e. the gauge and matter fields, are not inherently gauge invariant. Nonetheless, they are combined in a way such that the Lagrangian (or the Hamiltonian) itself is gauge invariant.
	A reformulation of these theories in terms of gauge invariant quantities allows a description purely in terms of physical degrees of freedom, even though the treatment can get more complicated from the computational point of view, as commented in \cite{Scwhartz} for example. Among several possible advantages, a crucial point is that this can be particularly helpful for the construction of consistent approximation schemes.
		
	To accomplish this, various attempts are present in literature, either on the lattice or in the continuum. 
	In the first case, scalar quantum electrodynamics (QED) and $SU(2)$ LGT in presence of bosonic matter fields were investigated in \cite{lattice_kijowski}. The main idea consists in introducing the gauge invariants of the corresponding continuum theories to rewrite the Lagrangians and derive the associated dynamics. In the continuum, this change of variables was applied to bosonic matter fields \cite{KIJOWSKI_nonabelianHiggs} and later to classical \cite{gaugeinv_rudolph_1} and quantum  \cite{gaugeinv_rudolph_2, gaugeinv_rudolph_4} electrodynamics.  In particular, the matter fields are combined into bosonic fields. In the case of $1+1$ dimensions, the Schwinger model, it was shown \cite{gaugeinv_rudolph_3} that the construction is related to the bosonization of the original theory (see \cite{COLEMAN1,COLEMAN2}).  
	More recently, $SU(2)$ LGT with fundamental fermions were studied and reformulated via the so-called loop-string-hadron formulation \cite{Stryker}: this allows for a description of the dynamics of the theory in terms of local and physical observables, using strictly $SU(2)$ gauge invariant variables at the cost of introducing extra lattice links and an Abelian Gauss law. In \cite{Kaplan-Stryker,Haase-Dellantonio,Bender-Zohar} the problem is addressed making use of dual formulations for the case of $U\left(1\right)$ gauge symmetry and having as a particular motivation the implementation of gauge theories in quantum devices. 
	
	The aim of the present paper is to set up a formalism allowing to reformulate abelian gauge theories in terms of gauge invariant fields (GIF). 
	%This construction is derived for abelian gauge theories on the lattice. 
	We look for a reformulation satisfying the following requirements: {\it i)} it should naturally extend to continuum theories; {\it ii)} it should allow to investigate the dependence on the particular construction used to eliminate the gauge covariant quantities; {\it iii)} in presence of matter fields (generically denoted with $\psi$), it should allow to straightforwardly determine a gauge invariant combination (denoted with $\psi'$) of the original matter and gauge variables.
	
	Fulfilling the last requirement {\it iii)} is the key point of the formalism presented here, since once the GIF $\psi'$ has been constructed it can be used as a new gauge invariant degree of freedom of the theory. While generic expectation values of functions of $\psi$ are not gauge invariant, the corresponding expectation values of $\psi'$ are fully physical. This is particularly relevant for finding suitable order parameters in light of Elitzur's theorem, which states that gauge symmetry cannot be spontaneously broken \cite{Elitzur}. For this reason, we think that the present formalism could be useful to analyze the phase diagram of gauge theories using approximate methods, such as the mean-field one, since it gives information on possible -- gauge invariant -- order parameters of the theory.
	
	We observe that the choice of the GIF is not unique: other combinations of the initial gauge and matter covariant operators can be gauge invariant. 
	A choice for the definition of $\psi'$ should emerge from the procedure. Our procedure leads to a simple expression for $\psi'$ in the form $\psi' \sim {\cal E} \psi$, where ${\cal E}$ is an operator depending only on the gauge field. Once the choice has been done, one can perform the elimination of the initial matter and gauge fields with different geometrical constructions, whose role is explicitly discussed in the following Sections. This structure of the GIF is analogous to the one already introduced by Dirac in \cite{Dirac1955}, where the proper gauge invariant operator creates the electron along with a "photon cloud" around it. A similar structure is also present in \cite{Gaete1997}, where gauge invariance is traded by a path-dependent choice of gauge invariant variables.
	
	Regarding {\it ii)}, we present two different constructions to split the gauge field into its gauge invariant and gauge covariant part. The gauge invariant part is taken as a new variable, while the gauge covariant part can be combined with matter to obtain, as anticipated, a new GIF. These constructions are presented -- keeping the size of the system finite -- both on the lattice and in the continuum for arbitrary dimension and for two kinds of boundary conditions: periodic and open. Finally, to show the effects that different constructions may have on physical models, we consider the Hofstadter model in $2d$ and $3d$, rewriting the Hamiltonian using both the procedures. This modifies the structure of the problem in momentum space, entering the definition of the so-called magnetic Brillouin zone (MBZ).
	
	The paper is organized as follows. In Sec. \ref{gauge_invariance} we briefly remind the concept of abelian gauge invariance in field theory,
	defining the various quantities and
	explaining the main ideas behind our constructions. In Sec. \ref{asymm_lattice_construction} we present our
	construction for a $(1+1)$ dimensional square lattice and discuss how to properly enforce open (OBC) and periodic (PBC) boundary conditions. 
	In Sec. \ref{symm_lattice_construction} we present
	a second construction for the $(1+1)$ dimensional case, alongside the one showed in Sec. \ref{asymm_lattice_construction}.
	In Sec. \ref{high_dims} we extend the reformulation to higher dimensions,
	firstly on the lattice and then taking the continuum limit. In Sec. \ref{latt_rewriting} we show how to reconstruct the lattice action for pure abelian gauge theories on the lattice, focusing on the particular example of gauge theories in $(2+1)$ dimensions. In Sec. \ref{qed_rewriting} we apply the formalism to the Hamiltonian of a single particle in a magnetic field and to the Lagrangian of QED. In Sec. \ref{hofstadter_rewriting} we consider the Hofstadter Hamiltonian in $2d$ and $3d$, showing how they are written in terms of the new gauge invariant variables. In Sec. \ref{discussion_applications} we comment about the applications of the reformulation. In Sec. \ref{conclusions} we summarize our results and present our conclusions.
	
	\section{\label{gauge_invariance}Abelian gauge theories}
	In quantum field theory, a gauge theory is generically described by a Lagrangian $\mathcal{L}[\psi,\bar{\psi},A_\mu]$, depending on some matter fields $\psi$, $\bar{\psi}$ and on a gauge field $A_\mu$. The theory is gauge invariant if the above Lagrangian does not change under local transformations $G(x)\in\mathcal{G}$, where $\mathcal{G}$ is the gauge group of the theory \cite{Peskin,Scwhartz,Maggiore}. From now on we will consider the specific case of abelian groups, referring to the abelian gauge theories. Formally, the previous local transformations can be written as
	\begin{equation}
	\psi(x)\rightarrow G(x)\psi(x),\hspace{1cm}\overline{\psi}(x)\rightarrow\overline{\psi}(x)G^{-1}(x)
	\label{gaugetransf_continuum_matterfield}
	\end{equation}
	for the matter fields and
	\begin{equation}
	A_\mu\rightarrow A_\mu-\frac{i}{q}[\partial_\mu G(x)] G^{-1}(x)
	\label{gaugetransf_continuum_gaugefield}
	\end{equation}
	for the gauge field. The field strength tensor $F_{\mu\nu}$ takes the form
	\begin{equation}
	F_{\mu\nu}\equiv\partial_\mu A_\nu-\partial_\nu A_\mu.
	\end{equation}
	Under a gauge transformation, this quantity is left unchanged, i.e. $F_{\mu\nu}\rightarrow F_{\mu\nu}$, therefore it represents a gauge invariant of the theory. For $(d+1)$ dimensions the indices $\mu,\nu\in\{0,\ldots,d\}$ where the index $0$ represents time.
	
	A prominent example of an abelian gauge theory is QED. The Lagrangian reads
	\begin{equation}
	\mathcal{L}=\bar{\psi}(i\slashed{\partial}-m)\psi-eA_\mu\bar{\psi}\gamma^\mu\psi-\frac{1}{4}F_{\mu\nu}F^{\mu\nu}.
	\label{QED_lagrangian}
	\end{equation}
	where $\psi$ and $\bar{\psi}$ are the fermionic degrees of freedom, $A_\mu$ is the gauge field and $m$ and $e$ are, respectively, the fermionic mass and charge parameters. The gauge group is $\mathcal{G}=U(1)$ and a generic local transformation can be written as a phase factor $G(x)=\exp\left(i e \Lambda(x)\right)$.	
	
	As mentioned in Sec. \ref{intro}, this kind of theories can be regularized on the lattice, following essentially two paths. The first one entails the discretization of the continuum theory Lagrangian. This constitutes the Lagrangian formalism of LGT.  There is also the possibility of considering the Hamiltonian formalism, in which space dimensions are discretized but time is not. In this formulation the theory is projected only on its physical states $|\Psi\rangle$, i.e. the ones satisfying Gauss's law \cite{Kogut-Susskind,Dalmonte-Montangero}.
	In these discretization schemes involving fermions, it is well known that particular attention must be paid to  address the fermion doubling problem \cite{Rothe}. This can be done by considering different discretizations of the fermionic field (e.g. Wilson fermions, staggered fermions or domain wall fermions). As these schemes preserve gauge invariance, our construction is largely independent on the type of fermions one uses on the lattice.
	
	\subsection{\label{basics_construction}Basic quantities on the lattice}
	The rewriting that will be presented can be defined both in the continuum and on the lattice, and, in the latter case, both in the Lagrangian and Hamiltonian formalisms. We will introduce the formalism on the lattice, in the Lagrangian formalism, where the path integral is mathematically well defined and the procedure more transparent. In order to make the exposition clear and fix the notation, we briefly review the basic ingredients for the standard discretization of a continuum gauge theory.
	
	A a generic site, on a $(d+1)$ dimensional lattice, is denoted by $(d+1)$ integers $\mathbf{n}=(n_0,\dots, n_d)$, where each component  takes values between $1$ and $N$. The gauge field $A_\mu$ is defined on the links of the lattice, while the field strength tensor $F_{\mu\nu}$ lives on the plaquettes. It is useful to define \cite{Rothe}
	\begin{equation}
	U_\mu(\mathbf{n})=e^{ieaA_\mu(\mathbf{n})},\hspace{0.8cm}U_{\mu\nu}(\mathbf{n})=e^{iea^2F_{\mu\nu}(\mathbf{n})},
	\end{equation}
	where $U_\mu\in\mathcal{G}$ are the link variables connecting the site $\mathbf{n}$ to the site $\mathbf{n}+\hat{\mu}$, for $\mu\in\{0,\dots,N\}$, and $U_{\mu\nu}(\mathbf{n})$ are the plaquette variables. The discretized version of the field strength tensor is written as
	\begin{equation}
	aF_{\mu\nu}(\mathbf{n})\equiv A_\nu(\mathbf{n}+\hat{\mu})-A_\nu(\mathbf{n})-A_\mu(\mathbf{n}+\hat{\nu})+A_\mu(\mathbf{n}).
	\label{discretized_Fmunu}
	\end{equation}
	The quantities $e$ and $a$ are respectively the charge and the lattice spacing. %From now on we will set $a=e=1$, and we will restore the lattice spacing to take the continuum limit. 
	We set $a=1$ and we will only recover it once we take the continuum limit.
	
	Referring explicitly to the $\mathcal{G}=U(1)$ gauge group, the action is given by
	\begin{equation}
	S=S_G[U_{\mu\nu}]+S_{\text{fermions}}[\psi,\bar{\psi},U_\mu],
	\end{equation}
	where
	\begin{equation}
	S_G=\frac{1}{e^2}\sum_P\bigg[1-\frac{1}{2}(U_{\mu\nu}+U^\dag_{\mu\nu})\bigg]
	\end{equation}
	is the pure gauge contribution, with the sum extended over all the plaquettes $P$, and $S_{\text{fermions}}$ represents the interaction with matter, whose explicit form depends on the discretization scheme used to treat the fermions. 
		
	In the Hamiltonian formalism we have again a pure gauge term plus interactions with matter
	\begin{equation}
	H_{\text{QED}}=H_G+H_{\text{fermions}}.
	\end{equation}
	Analogously, the latter depends on the explicit form of the fermion discretization,
	while the pure gauge part is given by
	\begin{equation}
	H_G=\frac{e^2}{2}\sum_{\mathbf{n},i}E^2_i(\mathbf{n})-\frac{1}{4e^2}\sum_P (U_{ij}+U^\dag_{ij}).
	\end{equation}
	The operators $E_i$ represent the electric field and are the canonically conjugate momenta of $A_i$, while $U_{ij}$ represent the magnetic field. 
	In the Hamiltonian formalism, gauge symmetry is manifested by the existence of a set of local generators that commute with the Hamiltonian
		\begin{equation}
		\left[G(\mathbf{n}),H_{\text{QED}}\right]=0,\quad \forall \mathbf{n}.
		\end{equation}
		The physical states are the ones that satisfy Gauss's law
		\begin{equation}
		G(\mathbf{n})|\Psi\rangle=0,\quad \forall \mathbf{n}
		\end{equation}
	where $G(\mathbf{n})=\nabla\cdot\mathbf{E}(\mathbf{n})-\rho(\mathbf{n})$ 
	and $\rho(\mathbf{n})$ is the charge density at the site $\mathbf{n}$.
	
	\subsection{\label{main_idea}Main idea}
	The main idea behind our formalism is to use eq. \eqref{discretized_Fmunu} to express the gauge field $A_\mu$ as a function of $F_{\mu\nu}$. 
	Clearly, this is not uniquely defined, since $F_{\mu\nu}$ is gauge invariant while $A_\mu$ is not. Put differently, $F_{\mu\nu}$ does not carry the gauge covariant part of $A_\mu$, which, instead, will be carried by a new field $\phi$. This will amount to replacing $A_\mu$ by a combination of $F_{\mu\nu}$ and $\phi$. In turn, not all the components of $F_{\mu\nu}$ are independent. The idea is then to define independent sums of $F_{\mu\nu}$ over various strips on the lattice, which will be denoted by  $\bar{F}_{\mu\nu}$. The presented formalism allows us to perform the change of variables
	\begin{equation}
	A_\mu\rightarrow\{\bar{F}_{\mu\nu},\phi\}.
	\label{main_mapping}
	\end{equation}
	We will show how to reconstruct the Lagrangian with the new variables in two ways, referred in the following as \textit{asymmetric} and \textit{symmetric} constructions. One has to check, for each construction, whether the defined $\bar{F}_{\mu\nu}$ are independent, and, if not, what are the relations between them. Our procedure bears some similarities with  the path-dependent choice of gauge invariant fields in \cite{Gaete1997}. In contrast to that approach, all of our gauge invariant variables are independent and do not have to satisfy any constraint.
		
	We emphasize that $\phi$ is a field and it is not fixed by our procedure: in the computation of the generating functional of the theory, we must sum over all the possible configurations of $\bar{F}_{\mu\nu}$. In turn, we can fix $\phi$, corresponding to choosing a specific gauge, or sum over it. The result will be the same. Contrary to standard gauge fixing, this will not alter the form of the Lagrangian, as the non-physical degrees of freedom were decoupled. This description holds both for OBC and PBC cases. However, in order to correctly reproduce all the degrees of freedom, some extra care is needed for the latter. To this end, a further new set of variables, associated with Wilson loops, will be introduced for PBC.
	
	We anticipate that when the matter fields $\psi$ are present, an advantage of our reformulation with respect to other possible ones is that the introduction of $\phi$ naturally indicates how to rewrite $\psi$ in terms of a new gauge invariant matter field $\psi'$, combining both $\psi$ and $\phi$. Once the integration over $\phi$ is performed, part of the contribution of the gauge field remains through $\bar{F}_{\mu\nu}$ and -- as will be clarified in the following -- the theory will be expressed in terms of the fields $\psi'$ and  $\bar{F}_{\mu\nu}$.
	
	\begin{figure}
		\centering
		\includegraphics[width=0.8\linewidth]{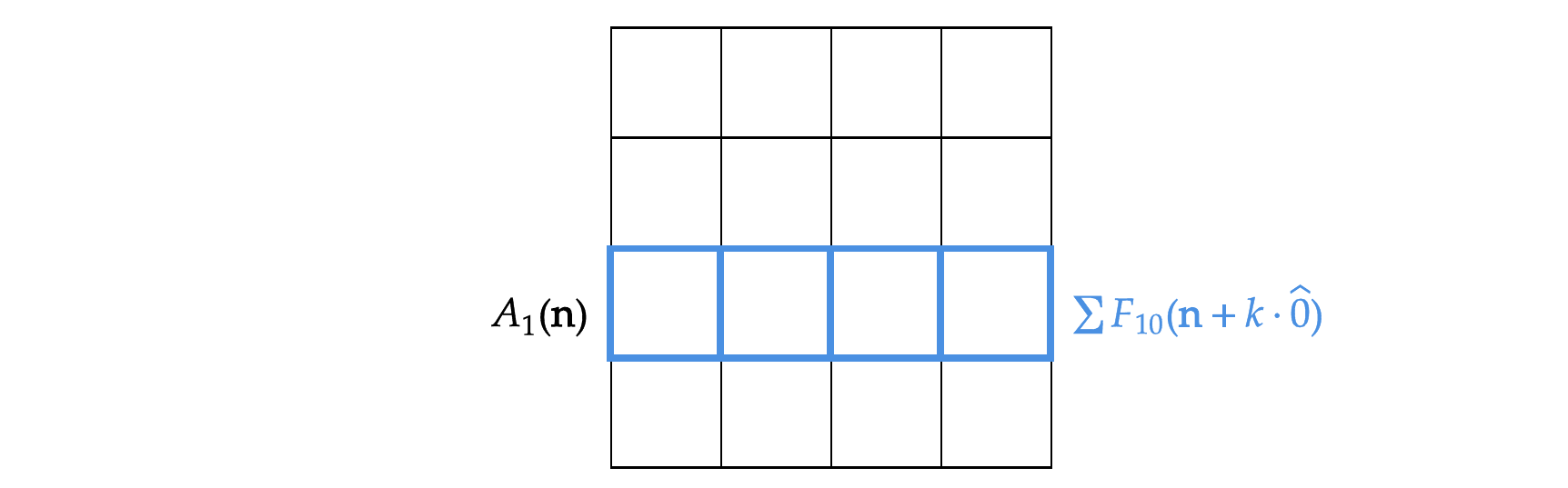}
		\caption{Iterative isolation of the link $A_1(\mathbf{n})$, the leftmost in the blue column, in terms of plaquettes. The highlighted strip  is the sum of $F_{10}$ present in Eq. \eqref{A1_beforevertices}. The temporal and horizontal direction are, respectively, the horizontal and vertical ones.}
		\label{strip_asymm}
	\end{figure}

	\section{\label{asymm_lattice_construction}Asymmetric construction}
\begin{figure}
	\centering
	\includegraphics[width=1.0\linewidth]{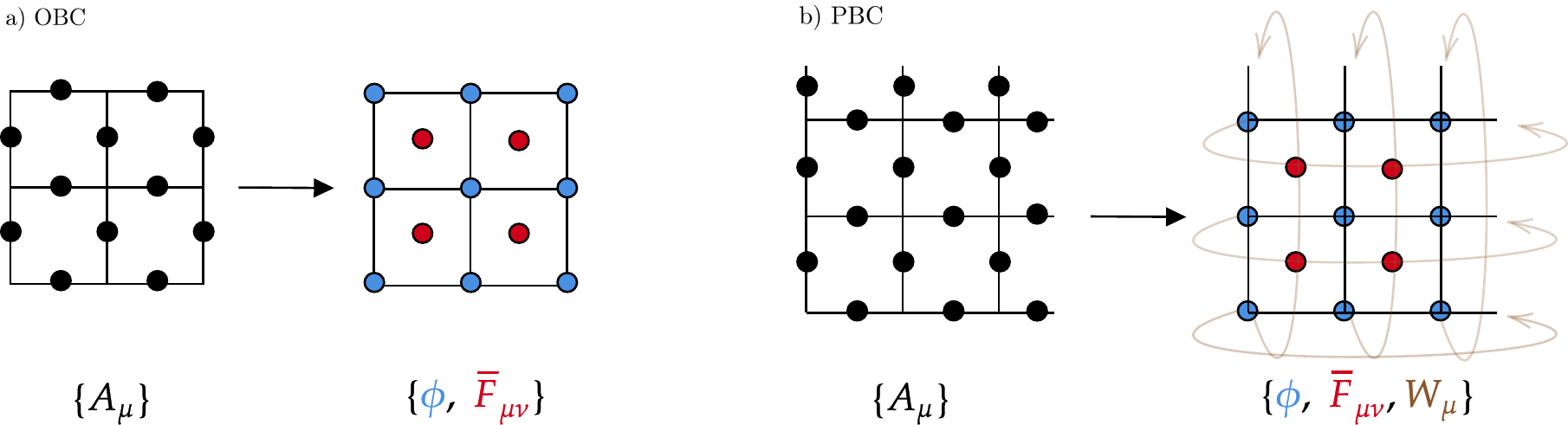}
	\caption{Plot of a lattice with linear size $N=3$. In black the original gauge field components $A_\mu$ living on the links. In color the representation of the new set of variables.  a) For OBC the new degrees of freedom are represented in blue (the $\phi$'s  that live on the vertices) and in red (the $\bar{F}_{\mu\nu}$'s  defined on the plaquettes). b) For PBC the same new degrees of freedom are present plus extra ones corresponding to the loops (the $f_\mu$ in brown).}
	\label{2d_changeofvars_picture}
	\end{figure}

	We start by presenting the asymmetric construction.
	In $(1+1)$ dimensions there is only one independent component of the strength tensor, i.e. $F_{10}$. Since we want to express the gauge fields $A_0$ and $A_1$ in terms of $F_{10}$, we use Eq. \eqref{discretized_Fmunu} to isolate  $A_1$ for a generic site $\mathbf{n}$
	\begin{equation}
	A_1(\mathbf{n})=F_{10}(\mathbf{n})-A_0(\mathbf{n}+\hat{1})+A_0(\mathbf{n})+A_1(\mathbf{n}+\hat{0}).
	\label{1step_A1}
	\end{equation}
	Graphically, this has a simple interpretation: it can be thought of as a plaquette with all the edges being removed except for $A_1(\mathbf{n})$.
	The next step is to use Eq. \eqref{1step_A1}, iteratively, to express all the $A_1$ links appearing on the right-hand side. This is done until the boundary $n_0=N$ is reached, as illustrated in Fig. \ref{strip_asymm}. In the end, we are left with the following expression for the gauge field:
	\begin{equation}
	A_1(\mathbf{n})=\sum_{k=0}^{N-n_0-1}\bigg[F_{10}(\mathbf{n}+k\cdot\hat{0})-A_0(\mathbf{n}+\hat{1}+k\cdot\hat{0})+A_0(\mathbf{n}+k\cdot\hat{0})\bigg]+A_1(\mathbf{n}+(N-n_0)\cdot\hat{0}).
	\label{A1_beforevertices}
	\end{equation}
	We now introduce the \textit{vertex variables} $\phi(\mathbf{n})$, defined on the vertices of the lattice, as shown in Fig. \ref{2d_changeofvars_picture}, and encoding the gauge covariant part of $A_\mu(\mathbf{n})$. Due to the nature of the considered gauge group $\mathcal{G}=U(1)$, these are scalar fields. The variables $\phi$ are related to $A_0$ through a finite derivative along the $\hat{0}$ direction, that is
	\begin{equation}
	A_0(\mathbf{n})\equiv\phi(\mathbf{n}+\hat{0})-\phi(\mathbf{n}).
	\label{A0_definition}
	\end{equation}
	Note that this can always be done. In turn, the choices for the values of $\phi$ are not unique, as we can always shift them by a function with arbitrary dependence on $n_1$ without changing the value of any $A_0\left(\mathbf{n}\right)$. This freedom will be explored in what follows.
	Inserting this into Eq. \eqref{A1_beforevertices} we obtain a telescopic sum, resolving the part associated to the horizontal links
	\begin{equation}
	A_1(\mathbf{n})=A_1(\mathbf{n}+(N-n_0)\cdot\hat{0})-[\phi(\mathbf{n}+(N-n_0)\cdot\hat{0}+\hat{1})-\phi(\mathbf{n}+(N-n_0)\cdot\hat{0})]+\phi(\mathbf{n}+\hat{1})-\phi(\mathbf{n})+\sum_{k=0}^{N-n_0-1}F_{10}(\mathbf{n}+k\cdot\hat{0}).
	\label{prima_di_BC}
	\end{equation}
	By exploiting the aforementioned freedom for choosing the field $\phi$, we can set
	\begin{equation}
	A_1(\mathbf{n}_B)=\phi(\mathbf{n}_B+\hat{1})-\phi(\mathbf{n}_B)
	\label{A1boundary_definition_OBC}
	\end{equation}		
	at the boundary points $\mathbf{n}_B\equiv(N,n_1)$. For the OBC case this essentially completes the map (see below), while for PBC further considerations are necessary.

	\subsection{\label{OBC_asymm_constr}Open boundary conditions}

	The OBC case is depicted in Fig. \ref{2d_changeofvars_picture} a). 
	By plugging Eq. \eqref{A1boundary_definition_OBC} into \eqref{prima_di_BC} we can write
	\begin{equation}
	A_1(\mathbf{n})=\sum_{k=0}^{N-n_0-1}F_{10}(\mathbf{n}+k\cdot\hat{0})+\phi(\mathbf{n}+\hat{1})-\phi(\mathbf{n})
	\label{A1_definition_OBC}
	\end{equation}
	which, with \eqref{A1boundary_definition_OBC}, concludes the rewriting of the vertical links. By defining the \textit{plaquette strip}
	\begin{equation}
	\bar{F}_{10}(\mathbf{n})\equiv\sum_{k=0}^{N-n_0-1}F_{10}(\mathbf{n}+k\cdot\hat{0}),
	\label{plaquettestrip_2d_OBC}
	\end{equation}
	the vertical links can be written as
	\begin{equation}
	A_1(\mathbf{n})=\bar{F}_{10}(\mathbf{n})+\phi(\mathbf{n}+\hat{1})-\phi(\mathbf{n}).
	\label{A1_definition_barF_OBC}
	\end{equation}
	We can characterize OBC by imposing $A_0(N,n_1)=0$ and $A_1(n_0,N)=0$. Consequently, Eqs. \eqref{A0_definition} and \eqref{A1_definition_barF_OBC} summarize the mapping of Eq. \eqref{main_mapping} for the OBC case, by adopting the boundary conditions $\bar{F}_{10}(N,n_1)=\bar{F}_{10}(n_0,N)=0$. There is yet the residual freedom on the choice of the field $\phi$. This is reflected by the fact that shifting all $\phi\left(\mathbf{n}\right)$ by a constant will leave the initial $A_\mu\left(\mathbf{n}\right)$ invariant. This ambiguity can be resolved by simply imposing $\phi\left(N,N\right)=0$, for example. We can now verify that the number of degrees of freedom matches the original one. There is a total of $N^2-1$ non-trivial values for $\phi$ and $(N-1)^2$ for $\bar{F}_{10}$, which sum to the original $2N(N-1)$ degrees of freedom associated with the links of open boundaries.
		
	\subsection{\label{PBC_asymm_constr}Periodic boundary conditions}
		
	The PBC case is depicted in Fig. \ref{2d_changeofvars_picture} b). 
	In comparison to OBC, we need to map an extra set of degrees of freedom. These correspond to links emanating from the boundary, i.e. $A_0(N,n_1)$ and $A_1(n_0,N)$, as well as to specify how the fields transform under a full lattice translation, i.e. $A_\mu(\mathbf{n}+N\hat{\nu})$. Regarding the first set of variables, we
	introduce the Wilson loops $W_{\mathcal{C}_0}$ and $W_{\mathcal{C}_1}$, which are associated to paths that wrap around the lattice along the $\hat{0}$ and $\hat{1}$ directions. We may formally write these loops as \cite{Rothe}
	\begin{equation}
		W_{\mathcal{C}_0}=\prod_{n_i=1}^Ne^{ieA_0(n_i,n_1)}\equiv e^{ief_0(n_1)},\qquad W_{\mathcal{C}_1}=\prod_{n_i=1}^N e^{ieA_1(n_0,n_i)}\equiv e^{ief_1(n_0)},
		\label{wilsonloops_defs}
	\end{equation}
	where $f_\mu(n_\nu)$ corresponds to the sum of all $A_\mu$ along a straight line of constant $n_\nu$ ($\mu\neq\nu$).
	Using the definitions in Eqs. \eqref{A0_definition}, \eqref{A1_definition_OBC} to rewrite the gauge fields for the remaining links, we can isolate the boundary fields as functions of the introduced loops as
	\begin{equation}
		A_0(N_,n_1)=f_0(n_1)-\phi(N,n_1)+\phi(1,n_1),\qquad A_1(n_0,N)=f_1(n_0)-\phi(n_0,N)+\phi(n_0,1)-\sum_{n_j=1}^{N-1}\bar{F}_{10}(n_0,n_j).
		\label{PBC_A0_A1_boundary_defs}
	\end{equation}
	By complementing Eqs. \eqref{A0_definition}, \eqref{A1boundary_definition_OBC} and \eqref{A1_definition_barF_OBC} with the above expression we can see that the mapping $\{A_0,A_1\}\rightarrow\{\phi,\bar{F}_{10},f_0,f_1\}$ is one-to-one. In particular, the number of degrees of freedom correctly match. In fact, the link variables $A_\mu$ form a total of $2N^2$ degrees of freedom. In turn the set $\{\phi,\bar{F}_{10}\}$ introduced in the OBC case has $2N(N-1)$ non-trivial values and the remaining $2N$ degrees of freedom are precisely given by the set of $2N$ loops $\{f_0,f_1\}$.
	It remains to specify how the fields $A_\mu(\mathbf{n}+N\hat{\nu})$ are expressed in terms of the gauge invariant degrees of freedom.
	As PBC should only be imposed on physical fields, the most general form of PBC on a gauge theory amounts to imposing periodicity on $A_\mu$ up to a gauge transformation \cite{THOOFT1979,mack1980sufficient,THOOFT1981,Wiese_habilitation}. Explicitly this means that
	\begin{equation}
	A_\mu(\mathbf{n}+N\hat{\nu})=A_\mu(\mathbf{n})+\varphi_\nu(\mathbf{n}+\hat{\mu})-\varphi_\nu(\mathbf{n}),
	\label{twist_bc}
	\end{equation}
	where $\varphi_\nu$ are called \textit{transition functions} and are crucial to study non-trivial topological sectors of the theory \cite{Wiese_habilitation,makeenko_2002}. They have to satisfy a consistency condition, called the \textit{cocycle condition}, guaranteeing that certain quantities, such as $A_\mu(\mathbf{n}+N\hat{0}+N\hat{1})$, are single valued. Such a condition reads
	\begin{equation}
		\varphi_\nu(\mathbf{n}+N\hat{\mu})+\varphi_\mu(\mathbf{n})=\varphi_\mu(\mathbf{n}+N\hat{\nu})+\varphi_\nu(\mathbf{n})+\varphi_{\nu\mu},
		\label{cocycle_cond}
	\end{equation}
	where $\varphi_{\mu\nu}$ is the \textit{twist tensor}, which is antisymmetric and gauge invariant \cite{Wiese_habilitation}. Moreover, the $\varphi_\nu$'s have to be considered as a set of new dynamical variables, i.e. physical degrees of freedom to be integrated in the functional integrals of the theory \cite{THOOFT1979}. 
	We can show that these boundary conditions can be incorporated within our reformulation.
	In fact, it follows from Eqs. \eqref{A0_definition}, \eqref{A1_definition_barF_OBC} and \eqref{PBC_A0_A1_boundary_defs} that the twisted boundary conditions in Eq. \eqref{twist_bc} are exactly implemented by the following boundary conditions on $\phi$
	\begin{equation}
	\phi(\mathbf{n}+N\hat{\nu})=\phi(\mathbf{n})+\varphi_\nu(\mathbf{n}).
	\label{twisted_bc_phi}
	\end{equation}
	The transition functions are the same in our reformulation and, consequently, the degrees of freedom that they carry are trivially translated to our construction. Concerning the strip variables $\bar{F}_{10}$, they are subject to PBC, i.e. $\bar{F}_{10}(\mathbf{n}+N\hat{\mu})=\bar{F}_{10}(\mathbf{n})$. This can be taken into account by appropriately redefining the strips as
	\begin{equation}
		\bar{F}_{10}(\mathbf{n})=\sum_{k=0}^{(N-n_0-1)\;\text{mod}\;N}F_{10}(\mathbf{n}+k\cdot\hat{0}).
	\end{equation}
	We finally observe that the loops $f_0,\;f_1$ obey PBC as long as periodic gauge transformations are considered. Indeed, the Wilson loops \eqref{wilsonloops_defs} acquire non-trivial phases under the application of topologically non-trivial gauge transformations, i.e. transition functions that are periodic up to integer multiples of $2\pi/e$ (sometimes such gauge transformations are called \textit{large} topologically non-trivial gauge transformations \cite{Wiese_habilitation}).
	This gauge redundancy can further be lifted by suitably combining the transition functions $\varphi_\nu$ with the loops $f_\nu$. Accordingly, we define
	\begin{equation}
		\bar{f_0}(n_1)=f_0(n_1)-\varphi_0(1,n_1),\qquad\bar{f_1}(n_0)=f_1(n_0)-\varphi_1(n_0,1).
	\end{equation}	
	The non-trivial phases acquired by the Wilson loops in Eq. \eqref{wilsonloops_defs} under large gauge transformations, corresponding to translations of the $f_\mu$, are canceled by the respective gauge transformations of the transition functions and we obtain quantities that are invariant under general gauge transformations. Under a full lattice translation, these loops are transformed by the twist tensor: $\bar{f}_0(n_1+N)=\bar{f}_0(n_1)+\varphi_{01}$ and $\bar{f}_1(n_0+N)=\bar{f}_1(n_0)+\varphi_{10}$.
			
	\subsection{Comments about gauge fixing}
	We showed that the mapping presented is defined in a consistent way, as any gauge field $A_\mu(\mathbf{n})$ can be expressed as a function of the new, independent, variables $\{\phi,\bar{F}_{10}\}$ through Eqs. \eqref{A0_definition}, \eqref{A1boundary_definition_OBC} and \eqref{A1_definition_barF_OBC} for OBC -- or as a function of  $\{\phi,\bar{F}_{10},f_0,f_1\}$ through  Eqs. \eqref{A0_definition}, \eqref{A1_definition_barF_OBC}, \eqref{A1boundary_definition_OBC} and \eqref{PBC_A0_A1_boundary_defs} for PBC. The price to pay is hidden in locality and translational invariance. The reformulation, as anticipated, shifts the effect of gauge transformations ${G}$ to the vertex variables, since $A_\mu\;\sim\;A_\mu+{\Lambda(\mathbf{n}+\hat{\mu})-\Lambda(\mathbf{n})}$, which is the lattice version of Eq. \eqref{gaugetransf_continuum_gaugefield}, implies $\phi(\mathbf{n})\sim\phi(\mathbf{n})+{\Lambda(\mathbf{n})}$.
	
	We observe that fixing a particular configuration of the vertex variables, e.g. $\phi=\tilde{\phi}$, would be equivalent to some gauge fixing $\cal{F}$, where one sums only over configurations that satisfy the constraint ${\cal F}\left(A_{\mu}\right)=0$. As an example one could choose $\tilde{\phi}=0$, which corresponds to a maximal tree gauge \cite{Creutz1977}.
	
	A careful reader could object that there is no real difference between our reformulation and other approaches corresponding to particular choices of $\phi$. However, the substantial difference is {\em not} that in these approaches $\phi$ is fixed and in ours is not, since the $\phi$ at the end will be anyway integrated out. At variance, the main difference will be clear when the matter field $\psi$ will be introduced: we will show that our reformulation allows for the definition of new matter variables $\psi'$ expressed in terms of the
	original variables $\psi$ and the to-be-integrated variables $\phi$. So the initial fields are $(\psi,A_\mu)$, which are separately gauge variants, while at the end of our reformulation procedure the theory is expressed in terms of the variables $(\psi',\bar{F}_{\mu\nu})$:
	\begin{equation}
	(\psi,A_\mu) \rightarrow (\psi',\bar{F}_{\mu\nu}),
	\label{main_mapping_matter}
	\end{equation}
	where the fields $(\psi',\bar{F}_{\mu\nu})$ are gauge invariant. In the case of PBC, new degrees of freedom are present through transition functions. Analogously, a reformulation in terms of purely gauge invariant fields can be achieved through $(\psi,A_\mu,\varphi_\mu) \rightarrow (\psi',\bar{F}_{\mu\nu},\bar{f}_\mu)$.
	
	\section{\label{symm_lattice_construction}Symmetric construction}
	Here we present an alternative construction starting from the set $\{\phi,F_{\mu\nu}\}$.
	As previously mentioned, if we choose to isolate the  temporal component in Eq. (\ref{1step_A1}) rather than the spatial one, we get vertical strips instead of the horizontal ones of the previous setup. The idea of the symmetric constructions is to remove such arbitrariness in the procedure and to combine both these asymmetric constructions, to obtain a more symmetric result.
	
	We proceed following the same structure of Sec. \ref{asymm_lattice_construction}, i.e. in $(1+1)$ dimensions. For the OBC case, the gauge fields in the asymmetric construction are written in Eqs. \eqref{A0_definition} and \eqref{A1_definition_barF_OBC} in terms of ${\bar{F}_{10}}$. However, if we had chosen $F_{01}$, the final formulas would have been 
	\begin{equation}
	A_1(\mathbf{n})\equiv\phi'(\mathbf{n}+\hat{1})-\phi'(\mathbf{n}),
	\label{A0_horiz_asymm}
	\end{equation}
	\begin{equation}
	A_0(\mathbf{n})=\sum_{k=0}^{N-n_1-1}F_{01}(\mathbf{n}+k\cdot\hat{1})+\phi'(\mathbf{n}+\hat{0})-\phi'(\mathbf{n})\equiv\bar{F}_{01}+\phi'(\mathbf{n}+\hat{0})-\phi'(\mathbf{n}).
	\label{A1_horiz_asymm}
	\end{equation}
	Here the fundamental variables are $\{\phi',\bar{F}_{01}\}$. To obtain a symmetric construction, we can define
	\begin{equation}
	\tilde{\phi}=\frac{\phi+\phi'}{2}
	\end{equation}
	and sum the previous relations with Eqs. \eqref{A0_definition}, \eqref{A1_definition_OBC}. The symmetrized gauge fields are
	\begin{equation}
	A_0(\mathbf{n})=\tilde{\phi}(\mathbf{n}+\hat{0})-\tilde{\phi}(\mathbf{n})+\frac{1}{2}\sum_{k=0}^{N-n_1-1}F_{01}(\mathbf{n}+k\cdot\hat{1}),
	\label{A0_symm_lattice}
	\end{equation}
	\begin{equation}
	A_1(\mathbf{n})=\tilde{\phi}(\mathbf{n}+\hat{1})-\tilde{\phi}(\mathbf{n})+\frac{1}{2}\sum_{k=0}^{N-n_0-1}F_{10}(\mathbf{n}+k\cdot\hat{0}).
	\label{A1_symm_lattice}
	\end{equation}
	This result can be obtained from the asymmetric construction, as in Eqs. \eqref{A0_horiz_asymm} and \eqref{A1_horiz_asymm}, by means of the gauge transformation
	\begin{equation}
	\phi\;\longrightarrow\;\tilde{\phi}-\frac{1}{2}\sum_{k=0}^{N-n_0-1}\sum_{\ell=0}^{N-n_1-1}F_{01}(\mathbf{n}+k\cdot\hat{0}+\ell\cdot\hat{1}).
	\label{gauge_transf}
	\end{equation}
	
	The specific details pertaining to PBC trivially extend to the symmetric construction. In particular, the boundary condition in Eq. \eqref{twisted_bc_phi} of the periodic case still holds with $\tilde{\phi}$ in place of $\phi$ and the boundary links in Eq. \eqref{PBC_A0_A1_boundary_defs} are symmetrized with respect to the strips $\bar{F}_{10}$ and $\bar{F}_{01}$, i.e.
	\begin{equation}
		A_0(N_,n_1)=f_0(n_1)-\tilde{\phi}(N,n_1)+\tilde{\phi}(1,n_1)-\frac{1}{2}\sum_{n_j=1}^{N-1}\bar{F}_{01}(n_j,n_1),
		\label{symm_PBC_boundaryA0}
	\end{equation}
	\begin{equation}
		A_1(n_0,N)=f_1(n_0)-\tilde{\phi}(n_0,N)+\tilde{\phi}(n_0,1)-\frac{1}{2}\sum_{n_j=1}^{N-1}\bar{F}_{10}(n_0,n_j).
		\label{symm_PBC_boundaryA1}
	\end{equation}
	where the loops $f_\mu$ are introduced according to the definition in Eq. \eqref{wilsonloops_defs}. Finally we notice that, within this particular construction, the strips $\bar{F}_{10}$ and $\bar{F}_{01}$ satisfy the relation $\bar{F}_{01}(\mathbf{n})+\bar{F}_{10}(\mathbf{n}+\hat{0})-\bar{F}_{10}(\mathbf{n})-\bar{F}_{10}(\mathbf{n}+\hat{1})=0$, 
	which shows explicitly that they are not all independent variables.

	\section{\label{high_dims}Higher dimensions and the continuum limit}
	We now consider $(d+1)$ dimensions, generalizing the previous construction to a hypercubic lattice of size $N^{d+1}$.
	We choose a reference plane to which we apply the procedure described in Sec. \ref{asymm_lattice_construction}. Then, given an arbitrary link, we can repeatedly apply identities of the form of Eq. \eqref{1step_A1} until we arrive at the reference plane. We will choose the reference plane to be the $2d$ surface defined by $n_i=N$ for $i=2,\ldots,d$.
	
	To lighten the notation, we will denote a boundary site by
	\begin{equation}
	\mathbf{n}^{(\mu)}\equiv(\underbrace{N,\ldots,N}_{\mu+1-\text{times}},n_{\mu+1},\ldots,n_d).
	\end{equation}
	Accordingly, $\mathbf{n}^{(0)}$ represents a point at the boundary $0$, $\mathbf{n}^{(1)}$ a point at boundaries $0$ and $1$, and so on.
	Moreover, we generalize the plaquette strip on the lattice as
	\begin{equation}
	\bar{F}_{\mu\nu}(\mathbf{n}^{(\nu-1)})\equiv\sum_{\ell=0}^{N-n_\nu-1}F_{\mu\nu}(\mathbf{n}^{(\nu-1)}+\ell\cdot\hat{\nu}).
	\label{plaquettestrip}
	\end{equation}
	
	In this compact notation the rewriting of a generic component $A_\mu(\mathbf{n})$ of the gauge field  is
	\begin{equation}
	A_\mu(\mathbf{n})=\sum_{\nu<\mu}\bar{F}_{\mu\nu}(\mathbf{n}^{(\nu-1)})+\phi(\mathbf{n}+\hat{\mu})-\phi(\mathbf{n}).
	\label{compact_gaugefield_rewriting}
	\end{equation}
	As before, there is a component written solely in terms of vertex variables, i.e. $A_0(\mathbf{n})$, and all the others are built by filling the lattice with the plaquette strips. Now $A_0(\mathbf{n})$ fixes $\phi$ up to arbitrary translations by functions dependent on $n_1,\dots, n_d$ (but not $n_0$), a freedom that is explored to fix the remaining $A_i$ at the boundaries. In agreement with \eqref{compact_gaugefield_rewriting}, we can take $\bar{F}_{\mu\nu}(\mathbf{n}^{(j-1)})$, with $\mu>\nu$, as the new set of independent variables.
	
	\subsection{Open boundary conditions}
	As in the $(1+1)$ dimensional case, the considerations above are enough to establish the transformation to the new variables for OBC. Once again, the values of the field $\phi$ is completely fixed up to an overall shift by a constant which is used to set $\phi(N,\dots,N)=0$. Furthermore the plaquette strips, at the proper boundary, are put to zero as well
	\begin{equation}
		\bar{F}_{\mu\nu}(\mathbf{n}^{(\nu-1)})=0,\ \ n_\mu=N\ \mathrm{or}\ n_\nu=N.
	\end{equation}
	Here, we also observe that the degrees of freedom are properly matched. There is a total of $N^{d+1}-1$ vertex variables $\phi$. The strips $\bar{F}_{ij}$ entail $N^{d-1-j}(N-1)^2$ for any $0\leq j<i\leq d$, giving a total of $dN^{d+1}-(d+1)N^d+1$ strip variables. Summing these together we find the required $(d+1)N^d(N-1)$ link variables of the initial formulation.
	
	\subsection{\label{PBC_higherdims}Periodic boundary conditions}	
	For the periodic case, the rewriting of $(1+1)$ dimensions also extends to higher dimensions, including the introduction of the loops $f_\mu$ defined in Eq. \eqref{wilsonloops_defs} and the corresponding gauge invariant $\bar{f}_\mu$. In particular, a generic boundary link $A_\mu(n_0,\ldots,n_\mu=N,\ldots,n_d)$ is expressed as
	\begin{align}
		\nonumber
		A_\mu(n_0,\ldots,n_\mu=N,\ldots,n_d)&=f_\mu(n_0,\ldots,n_{\mu-1},n_{\mu+1},\ldots,n_d)-\sum_{\nu<\mu}\sum_{n_\mu=1}^{N-1}\bar{F}_{\mu\nu}(\mathbf{n}^{(\nu-1)})\\
		&-\phi(n_0,\ldots,n_\mu=N,\ldots,n_d)+\phi(n_0,\ldots,n_\mu=1,\ldots,n_d).
		\label{boundarylinks_higherdim}
	\end{align}
	The transformation in Eq. \eqref{twisted_bc_phi} implements the periodicity on the gauge field of Eq. \eqref{compact_gaugefield_rewriting}, up to a gauge transformation in exactly the same way. Once again the $\bar{F}_{\mu\nu}$ are periodic and $\bar{f}_{\mu}$ transform with the twist tensor, i.e. $\bar{F}_{\mu\nu}(\mathbf{n}+N\hat{\delta})=\bar{F}_{\mu\nu}(\mathbf{n})$ and $\bar{f}_{\mu}(\mathbf{n}+N\hat{\delta})=\bar{f}_{\mu}(\mathbf{n})+\varphi_{\mu\delta}$.
	The number of degrees of freedom can be computed by summing the ones from OBC $(d+1)N^d(N-1)$ with the number of loops $f_\mu$ introduced $(d+1)N^d$. This gives $(d+1)N^{d+1}$, which matches exactly the number of starting links $A_\mu$.
	
	\subsection{\label{cont_lim_high_dims}Continuum limit}
	So far we developed the formalism on the lattice. It is straightforward  to
	%,but now we 
	take the continuum limit of Eq. \eqref{compact_gaugefield_rewriting}. 
	We recover the lattice spacing $a$ and take the limit $a\rightarrow0$ and $N\rightarrow\infty$ while keeping $Na\equiv L$ fixed. We obtain
	\begin{equation}
	\bar{F}_{\mu\nu}(\mathbf{x}^{(\nu-1)})=\int_{x_\nu}^{L}\;dy_\nu\;F_{\mu\nu}(\mathbf{y}_\nu)
	\label{continuum_strip}
	\end{equation}
	as the continuum counterpart of the plaquette strip, while the gauge field is written as
	\begin{equation}
	A_\mu(\mathbf{x})=\partial_\mu\phi(\mathbf{x})+\sum_{\nu<\mu}\bar{F}_{\mu\nu}(\mathbf{x}^{(\nu-1)}).
	\label{continuum_gaugefield}
	\end{equation}
	In the previous expressions we have introduced
	\begin{equation}
	\mathbf{x}^{(\nu)}\equiv(\underbrace{L,\ldots,L}_{\nu+1-\text{times}},x_{\nu+1},\ldots,x_d),
	\end{equation}
	\begin{equation}
	\mathbf{y}_\nu\equiv(\underbrace{L,\ldots,L}_{\nu-\text{times}},y_\nu,x_{\nu+1},\ldots,x_d),
	\end{equation}
	as a shorthand notation for the real space vectors, being $\nu\in\{0,\ldots,d\}$. We remark that this completely characterizes the case of open but not of periodic boundaries, since the lattice description of the latter relied on the special mapping of a single link, which does not generalize straightforwardly in the continuum limit at finite size $L$. We do not see conceptual problems in doing it, and we leave the explicit implementation of PBC at finite $L$ in the continuum limit as a subject for a future work.

	\section{\label{latt_rewriting}	Pure abelian gauge theories on the lattice}
	Before discussing systems with matter fields present, we provide a more concrete example. We consider the standard action for non-compact gauge fields, in imaginary time and in $(d+1)$ dimensions
	\begin{equation}
	S=\frac{\beta}{2} \sum_{\mathbf{n}} F_{\mu \nu}(\mathbf{n})^2
	\label{latt_gaugeS}
	\end{equation}
	where the sum is taken over the $N^{d+1}$ lattice points.
	The discussion can be translated for any action depending solely on $F_{\mu \nu}$. The main premise of the present paper is to rewrite the model purely in terms of gauge invariant quantities. This is already done in Eq. \eqref{latt_gaugeS} where the action only depends on $F_{\mu \nu}(\mathbf{n})$: however, \textit{they are not all independent}. They satisfy the Bianchi identity in the continuum, and, on the lattice, a discretized version that in this case can be written as
	\begin{equation}
	F_{\mu \nu}(\mathbf{n}+\hat{\alpha})-F_{\mu \nu}(\mathbf{n})+F_{\alpha \mu}(\mathbf{n}+\hat{\nu})-F_{\alpha \mu}(\mathbf{n})+F_{\nu \alpha}(\mathbf{n}+\hat{\mu})-F_{\nu \alpha}(\mathbf{n})=0.
	\label{bianchi}
	\end{equation}
	This identity is trivially satisfied when the $F_{\mu \nu}$ are written in terms of the $A_\alpha$. In other words, while $F_{\mu \nu}$ are gauge invariant but not all independent, $A_\alpha$ are gauge covariant but independent. With our construction we are able to achieve both gauge invariance and independence on the new variables. In fact, since the description of the theory in terms of $A_\alpha$ satisfies the Bianchi identity and Eq.  \eqref{compact_gaugefield_rewriting} is a rewriting of them in terms of independent quantities, the Bianchi identity will be automatically satisfied for this case. 
	
	In order to make the discussion even more specific, let us focus on the non-trivial case of the $(2+1)$ gauge theory. Following \eqref{compact_gaugefield_rewriting} we write
	\begin{align}
		& A_0(\mathbf{n})=\phi(\mathbf{n}+\hat{0})-\phi(\mathbf{n}),
		\nonumber
		\\& A_1(\mathbf{n})=\bar{F}_{10}(\mathbf{n})+\phi(\mathbf{n}+\hat{1})-\phi(\mathbf{n}),
		\nonumber
		\\& A_2(\mathbf{n})=\bar{F}_{20}(\mathbf{n})+\bar{F}_{21}(\mathbf{n^{(0)}})+\phi(\mathbf{n}+\hat{2})-\phi(\mathbf{n}).
		\label{A_3+1}
	\end{align}
	In these formulas $\bar{F}_{10}$ and $\bar{F}_{20}$ are defined in all lattice points. At variance, $\bar{F}_{21}$ is only defined for boundary points where $n_0=N$. 
	The plaquettes $F_{\mu\nu}(\mathbf{n})$ in terms of these fields are given by
	\begin{align}
		& F_{10}(\mathbf{n})=\bar{F}_{10}(\mathbf{n})-\bar{F}_{10}(\mathbf{n}+\hat{0}),
		\nonumber
		\\&F_{20}(\mathbf{n})=\bar{F}_{20}(\mathbf{n})-\bar{F}_{20}(\mathbf{n}+\hat{0}),
		\nonumber
		\\&F_{21}(\mathbf{n})=\bar{F}_{10}(\mathbf{n}+\hat{2})-\bar{F}_{10}(\mathbf{n})+\bar{F}_{20}(\mathbf{n})-\bar{F}_{20}(\mathbf{n}+\hat{1})+\bar{F}_{21}(\mathbf{n^{(0)}})-\bar{F}_{21}(\mathbf{n^{(0)}}+\hat{1}),
	\end{align}
	where we define any $\bar{F}_{10}$ to be zero outside of any point on the lattice.
	
	As a further check, one can see that the Bianchi identity is trivially satisfied once the $F_{\mu\nu}$ are written in this way.
	To emphasize the differences with the new description of the theory, let us denote $a_\mathbf{n} \equiv \bar{F}_{10}(\mathbf{n})$, $c_\mathbf{n} \equiv \bar{F}_{20}(\mathbf{n})$ and $b_{\mathbf{n}^{(0)}} \equiv \bar{F}_{21}(\mathbf{n^{(0)}})$ as a boundary field. We have, for a general lattice point, $\mathbf{n}=(n_0,n_1,n_2)$ and $(n_0,n_1,n_2)^{(0)}=(N,n_1,n_2)$. The action takes the form
	\begin{align}
		\nonumber
		S=&\beta \sum_{\mathbf{n}}
		\left[
		\left(a_{\mathbf{n}+\hat{0}}-a_{\mathbf{n}}\right)^2+
		\left(c_{\mathbf{n}+\hat{0}}-c_{\mathbf{n}}\right)^2+\left(a_{\mathbf{n}+\hat{2}}-a_{\mathbf{n}}\right)^2+
		\left(c_{\mathbf{n}+\hat{1}}-c_{\mathbf{n}}\right)^2
		\right]
		\\&\beta N \sum_{\mathbf{n},n_0=N}
		\left[
		\left(b_{\mathbf{\mathbf{n}}+\hat{1}}-b_{\mathbf{\mathbf{n}}}\right)^2\right]+2\beta \sum_{\mathbf{n}}
		\left[
		\left(b_{\mathbf{\mathbf{n}^{(0)}}+\hat{1}}-b_{\mathbf{\mathbf{n}^{(0)}}}\right)
		\left(a_{\mathbf{n}+\hat{2}}-a_{\mathbf{n}}
		-c_{\mathbf{n}+\hat{1}}+c_{\mathbf{n}}\right)
		\right].
		\label{latt_gaugeinv_S}
	\end{align}
	This is a non-isotropic, non-local model. The first two terms are purely local. The second term is a boundary term that, nonetheless, is not predicted to be negligible in the infinite volume limit since it has a prefactor $N$. The last term is non-local and couples fields in the bulk to the boundary fields (at the boundary $n_0=N$). The resulting non-locality can be regarded as the integration of the gauge covariant part of the gauge fields. Other examples of the integration of gauge fields lead naturally to non-local interactions \cite{hamer1997,barros2018long,barros2019string}. In contrast to the cited results, here the full gauge degrees of freedom are not totally integrated out but only their non-physical part.
	
	Despite the apparent complication of this model, it is described by fewer degrees of freedom, all of them physical. As an example, for the case of OBC where we counted the degrees of freedom, all the $N^3-1$ vertex variables $\phi$ have decoupled from the system.
	
	\section{\label{qed_rewriting}QED}
	In this Section we rewrite the Lagrangian of QED in terms of gauge invariant quantities, using the asymmetric construction in $(3+1)$ dimensions. 
 	We refer to the next Section, where we discuss the Hofstadter model, for a discussion of the effects 
	produced by the choice of the asymmetric \textit{vs} symmetric construction.
	In that example, the differences are particularly clear.
	Before dealing with QED case, we investigate the simple and instructive case of the Hamiltonian of a particle in the presence of an external magnetic field, in $3d$.
	
	\subsection{\label{landau_levels}Single particle in a magnetic field}
	We consider a quantum particle in a static backgroud magnetic field. As it is usually done in quantum mechanics textbooks, the particle is charged (with charge $-e$) and we denote the size of the system by $L$, taking then the thermodynamic limit $L\rightarrow\infty$. 
	The Hamiltonian reads
	\begin{equation}
	\mathcal{H}=\frac{(\mathbf{p}+e\mathbf{A})^2}{2m},
	\end{equation}
	where $\mathbf{p}=-i\nabla$ and we make use of natural units $\hbar=c=1$. The only non-trivial component of the field strength tensor is $F_{21}=B$, therefore we only have the strip $\bar{F}_{21}$ in the asymmetric construction. 
	This results in
	\begin{equation}
	A_i=\partial_i\phi,\hspace{0.5cm}i\neq2,
	\end{equation}
	\begin{equation}
	A_2=\partial_2\phi+\bar{F}_{21}=\partial_2\phi+B(L-x)
	\end{equation}
	where $i=1,2,3$. The crucial point consists in transforming the wavefunction, $|\psi\rangle$, and observables in such a way that we deal only with gauge invariant quantities, independent of $\phi$. This is achived by the unitary transformation $|\psi'\rangle=\exp\left(-ie\phi\right)|\psi\rangle\equiv S|\psi\rangle$ and $\mathcal{H}'=S\mathcal{H}S^{-1}$.	
	A closer look into the new momenta
	\begin{equation}
	p_i'=Sp_iS^{-1}=p_i+e\partial_i\phi
	\end{equation}
	confirms that the vertex variables are reabsorbed . It is easy to verify that both $|\psi'\rangle$ and $p_i'$ are gauge invariant.
	We conclude that
	\begin{equation}
	\mathcal{H}'=\frac{1}{2m}\sum_{i=1}^3\bigg(p_i'+\sum_{j<i}\bar{F}_{ij}\bigg)^2.
	\end{equation}
	is the correct rewriting for the Hamiltonian. Since the transformation $S$ is unitary, the spectra of $\mathcal{H}'$ and $\mathcal{H}$ coincide, reproducing the well-known Landau levels as expected \cite{Landau}.

	\subsection{\label{rewriting_QED}The QED Lagrangian}
	Let us now consider the QED Lagrangian of Eq. \eqref{QED_lagrangian}, defined in a cubic volume with OBC, for simplicity. It can be transformed into
	\begin{equation}
	\mathcal{L}=\bar{\psi}\bigg[i\slashed{\partial}-m-e(\partial_\mu\phi)\gamma^\mu-e\sum_{\nu<\mu}\bar{F}_{\mu\nu}\gamma^\mu\bigg]\psi-\frac{1}{4}\bigg[\sum_{\alpha<\nu}\partial_\mu\bar{F}_{\nu\alpha}-\sum_{\alpha<\mu}\partial_\nu\bar{F}_{\mu\alpha}\bigg]\bigg[\sum_{\alpha<\nu}\partial^\mu\bar{F}^{\nu\alpha}-\sum_{\alpha<\mu}\partial^\nu\bar{F}^{\mu\alpha}\bigg].
	\end{equation}
	using the continuum rewriting of the gauge field of Eq. \eqref{continuum_gaugefield}.
	Due to the presence of matter, this is not yet written in terms of gauge invariant fields alone. Consequently, we define
	\begin{equation}
	\psi'=\exp\left(ie\phi\right)\psi.
	\label{gauge_inv_newfields}
	\end{equation}
	The equation above is the most important result of this Section, since it provides 
	an expression of the GIF expressed in the form of the field operator $\psi$ of the initial fermionic operator, which is not gauge invariant, multiplied by an operator depending on the gauge degrees of freedom. Overall, $\psi'$ is gauge invariant.
	The term with the vertex variables is canceled from the Lagrangian, which finally reads
	\begin{equation}
	\mathcal{L}=\bar{\psi}'\bigg[i\slashed{\partial}-m-e\sum_{\nu<\mu}\bar{F}_{\mu\nu}\gamma^\mu\bigg]\psi'-\frac{1}{4}\bigg[\sum_{\alpha<\nu}\partial_\mu\bar{F}_{\nu\alpha}-\sum_{\alpha<\mu}\partial_\nu\bar{F}_{\mu\alpha}\bigg]\bigg[\sum_{\alpha<\nu}\partial^\mu\bar{F}^{\nu\alpha}-\sum_{\alpha<\mu}\partial^\nu\bar{F}^{\mu\alpha}\bigg].
	\label{QED_lagrangian_newvars}
	\end{equation}
	%\end{widetext}
	This completes the rewriting of the QED Lagrangian in terms of $\bar{F}_{\mu\nu}$ and the new fields $\psi',\;\bar{\psi}'$, which are combinations of the vertex variables and the original fermionic fields. When written explicitly we find a non-local structure of the Lagrangian both in the gauge kinetic part as well as in the coupling to the matter fields.
	In principle, it is possible to derive a Hamiltonian through canonical quantization. In practice, due to the non-locality of the kinetic term, this may be highly non-trivial
	\footnote{More precisely, writing explicitly the terms involving the strips in Eq. (\ref{QED_lagrangian_newvars}) we get that the only non-trivial conjugate momenta are associated to $\bar{F}_{i0},\;i=1,2,3$, the electric field components. The right way to proceed should be to introduce a set of Lagrange multipliers, associated to the vanishing conjugate momenta of the theory, i.e. $\bar{F}_{21,31,32}$. Once done that, it should be quantized as a constrained theory.}.
	
	\section{\label{hofstadter_rewriting}Hofstadter model}
	We present now the reformulation of the Hofstadter model in $2d$ and $3d$, in terms of the new variables. We do so by using the asymmetric and symmetric constructions, in order to discuss their differences and to show how they reproduce the correct results for the energy spectrum. 
	
	Such a model describes a non-relativistic particle hopping on a lattice under the effect of an external magnetic field $\mathbf{B}$. The Hamiltonian, assuming the Peierls substitution to take into account the effects of the external field \cite{Hofstadter}, is 
	\begin{equation}
	\mathcal{H}=-t\sum_{\mathbf{r},\hat{j}}c^\dag_{\mathbf{r}+\hat{j}}e^{i\theta_{\mathbf{r}+\hat{j},\mathbf{r}}}c_\mathbf{r}+\mathrm{h.c.},
	\label{hofstadter_hamiltonian}
	\end{equation}
	where $\hat{j}$ are unit vectors along the spatial directions of the lattice ($\hat{j}=\hat{x},\;\hat{y}$ in $2d$ and $\hat{j}=\hat{x},\;\hat{y},\;\hat{z}$ in $3d$) , $c^\dag_\mathbf{r},\;c_\mathbf{r}$ are the fermionic creation and annihilation operators and
	\begin{equation}
	\theta_{\mathbf{r}+\hat{j},\mathbf{r}}\equiv\int_{\mathbf{r}}^{\mathbf{r}+\hat{j}}\mathbf{A}(\mathbf{x})\cdot d\mathbf{x}
	\end{equation}
	is the Peierls phase. The vector potential $\mathbf{A}(\mathbf{x})$ is  associated with the external field. 
	In order to have an isotropic magnetic flux on each plaquette of the lattice, we consider a magnetic field whose magnitude is
	\begin{equation}
	\Phi=\frac{2\pi m}{n}
	\label{isotropic_field}
	\end{equation}
	where $m,\;n$ are coprime integer numbers. Its direction will be specified below. In the following, we consider cubic lattices with $V=N^d$ sites, $d$ being the dimension, and sizes $N=\kappa n$, with $\kappa\in\mathbb{N}$. The latter is a necessary condition to solve the problem analytically in momentum space, when PBC are imposed \cite{Burrello-Trombettoni1}.
	
	The $2d$ model was introduced in literature in \cite{Hofstadter,Harper} and it is a celebrated and paradigmatic model to study commensurability effects. In $3d$, its energy spectrum can be determined for generic pairs of coprimes $n$ and $m$: we refer to \cite{Burrello-Trombettoni1} for a review of the problem of diagonalizing the Hofstadter Hamiltonian for general $n$ and $m$ in $d=3$. There, it is also shown that for general $n$ and $m$ it is convenient to work in the so-called Hasegawa gauge, introduced in \cite{Hasegawa}. Here we rewrite the corresponding Hamiltonians using both the asymmetric and symmetric constructions \textcolor{black}{and not choosing any gauge}, showing how the obtained expressions lead to different structures in momentum space. 
		
	In the next two Sections we consider the $2d$ and $3d$ models, in both cases assuming PBC. Our aim is to explicitly show how the formal constructions presented in Sections \ref{PBC_asymm_constr}, \ref{symm_lattice_construction} and \ref{PBC_higherdims} work and reproduce the known results.
	
	\subsection{\label{2d}The $2d$ model}
	We consider a square lattice with $V=N^2$ sites and a perpendicular commensurate magnetic field $\mathbf{B}=\Phi(0,0,1)$. We change notation with respect to Sec. \ref{asymm_lattice_construction}, denoting a generic lattice site by $\mathbf{r}=(r_1,r_2)$, in order to avoid confusion with the integer $n$ appearing in Eq. \eqref{isotropic_field}.
	
	\subsubsection{Asymmetric construction}
	The only non-trivial component of the field strength tensor is $F_{21}$. We use then Eqs. \eqref{plaquettestrip} and \eqref{compact_gaugefield_rewriting} to rewrite the gauge field. The only non-zero plaquette strip is $\bar{F}_{21}=\Phi(N-r_1)$, with $r_1<N$. We have still to specify the gauge invariant loops $\bar{f}_i$: they can be determined by imposing that the flux on the plaquettes of the boundary sites $\mathbf{r}_{B,1}=(N,r_2)$, $\mathbf{r}_{B,2}=(r_1,N)$ is equal to $\Phi$, so we have a uniform magnetic field through the whole lattice. Using the definitions in Eq. \eqref{PBC_A0_A1_boundary_defs} we get
	\begin{equation}
		\bar{f}_1(r_2)=\Phi N r_2+\vartheta_1,\qquad \bar{f}_2(r_1)=-\Phi N r_1+\vartheta_2.
		\label{loops_2dhofst}
	\end{equation}
	The constants $\vartheta_1,\;\vartheta_2$ account for twists of the fermionic operators at the boundaries. They are gauge invariant physical quantities that should be specified along with the magnetic field. In order to compare our construction with the known results in the literature for PBC, we choose these parameter to be $\vartheta_1=\vartheta_2=0$ (for further discussion on these parameters we refer to \cite{al-wiese2009}).  The Hamiltonian \eqref{hofstadter_hamiltonian} is rewritten as
	\begin{equation}
		\nonumber
		\mathcal{H}=-t\sum_{\mathbf{r}\neq\mathbf{r}_{B,i}}(c_{\mathbf{r}+\hat{1}}^\dag e^{i[\phi(\mathbf{r}+\hat{1})-\phi(\mathbf{r})]} c_{\mathbf{r}}+c_{\mathbf{r}+\hat{2}}^\dag e^{i[\phi(\mathbf{r}+\hat{2})-\phi(\mathbf{r})+\bar{F}_{21}]} c_{\mathbf{r}}+\text{h.c.})+\mathcal{H}_B,
	\end{equation}
	where the boundary terms are
	\begin{equation}
		\mathcal{H}_B=-t(c^\dag_{\mathbf{r}_{B,1}+\hat{1}}e^{i(f_1(r_2)-\phi(N,r_2)+\phi(1,r_2))}c_{\mathbf{r}_{B,1}}+c^\dag_{\mathbf{r}_{B,2}+\hat{2}}e^{i(f_2(r_1)-\phi(r_1,N)+\phi(r_1,1)-(N-1)\bar{F}_{21}(r_1))}c_{\mathbf{r}_{B,2}}+\text{h.c.}).
	\end{equation}
	The fermionic operators at the boundaries transform as
	\begin{equation}
		c^\dag_{\mathbf{r}_{B,1}+\hat{1}}=e^{-i\varphi_1(1,r_2)}c^\dag_{(1,r_2)},\qquad c^\dag_{\mathbf{r}_{B,2}+\hat{2}}=e^{-i\varphi_2(r_1,1)}c^\dag_{(r_1,1)},
	\end{equation}
	which allow us to suitably identify gauge invariant loops $\bar{f}_i$ in the hopping phases of $\mathcal{H}_B$ and replace them with their values \eqref{loops_2dhofst}.
	Analogously to the QED case, we define new fermionic gauge invariant operators 
	\begin{equation}
		d_{\mathbf{r}}\equiv e^{-i\phi(\mathbf{r})}c_{\mathbf{r}},\hspace{1cm}d^\dag_{\mathbf{r}}\equiv c^\dag_{\mathbf{r}}e^{i\phi(\mathbf{r})}.
		\label{gauge_inv_fermions}
	\end{equation}
	The operator $d_{\mathbf{r}}$ is the equivalent of the GIF $\psi'$ introduced for QED in Eq.  \eqref{gauge_inv_newfields}. The gauge invariance of the operator $d_{\mathbf{r}}$, as well as its fermionic nature, is explicit. A gauge transformation of function $\Lambda(\mathbf{r})$ modifies the vertex variables through the shift $\phi(\mathbf{r})\sim\phi(\mathbf{r})+\Lambda(\mathbf{r})$, exactly canceled by the phases of the gauge transformed operators $c_\mathbf{r},\;c^\dag_\mathbf{r}$, see Eq. \eqref{gaugetransf_continuum_matterfield}.
	
	It is now immediate to check that the boundary terms in $\mathcal{H}_B$ have the same structure of the bulk terms. This is due to the definitions of $\Phi$, $N$ (since $\Phi N$ is an integer multiple of $2\pi$) and the chosen values of $\vartheta_1=\vartheta_2=0$. Indeed we have
	\begin{equation}
		e^{i\bar{f}_1(r_2)}=e^{i\Phi N r_2}=1,\qquad e^{i\bar{f}_2(r_1)}=e^{i(\Phi(N-N^2)-\Phi r_1)}=e^{-i\Phi r_1}
	\end{equation}
	As a consequence,
	the Hamiltonian in terms of the new gauge invariant variables is
	\begin{equation}
		\mathcal{H}=-t\sum_{\mathbf{r}}(d^\dag_{\mathbf{r}+\hat{2}}e^{-i\Phi r_1}d_{\mathbf{r}}+d^\dag_{\mathbf{r}+\hat{1}}d_{\mathbf{r}}+\mathrm{h.c.}).
	\end{equation}
	We remark that the above description of the physical system, which does not reference gauge covariant operators, was achieved without ever fixing a gauge.
	
	We now move to the momentum space, introducing the Fourier transformed operators
	\begin{equation}
		d_{\mathbf{r}}=\frac{1}{\sqrt{V}}\sum_\mathbf{k}d_{\mathbf{k}}e^{i\mathbf{k}\cdot\mathbf{r}},\hspace{0.5cm}d^\dag_{\mathbf{r}}=\frac{1}{\sqrt{V}}\sum_\mathbf{k}d^\dag_{\mathbf{k}}e^{-i\mathbf{k}\cdot\mathbf{r}}.
	\end{equation}
	The full Hamiltonian then becomes
	\begin{equation}
		\mathcal{H}=-t\sum_\mathbf{k}(2\cos{k_1}d^\dag_\mathbf{k}d_\mathbf{k}+e^{-ik_2}d^\dag_{\mathbf{k}+\Phi\hat{1}}d_{\mathbf{k}}+\mathrm{h.c.})
	\label{2d_tb_momentumspace}
	\end{equation}
	where the momenta are chosen in the first Brillouin zone (1BZ), i.e. the square $[-\pi,\pi)\times[-\pi,\pi)$.
	
	The interplay between gauge and translational invariance in presence of a commensurate background magnetic field allows us to introduce the concept of magnetic Brillouin zone  \cite{Lifschitz-Pitaevski}. In this case, it is given by
	\begin{equation}
		\mathrm{MBZ:}\hspace{0.5cm}k_1\in\bigg[-\frac{\pi}{n},\frac{\pi}{n}\bigg),\;k_2\in\bigg[-\pi,\pi\bigg).
		\label{2d_asymm_MBZ}
	\end{equation}
	This enables us to split the structure of the Hamiltonian in terms of the so-called magnetic bands, labeled by an index $\tau\in\{0,1,\ldots,n-1\}$:
	\begin{equation}
		\mathcal{H}=-t\sum_{\mathbf{k}\in\mathrm{MBZ}}\sum_\tau[2\cos{(k_1+\tau \Phi)}d^\dag_{\mathbf{k}+\tau \Phi\hat{1}}d_{\mathbf{k}+\tau \Phi\hat{1}}+e^{-ik_2}d^\dag_{\mathbf{k}+(\tau+1) \Phi\hat{1}}d_{\mathbf{k}+\tau \Phi\hat{1}}+\mathrm{h.c.}].
		\label{2d_symm_hamiltonian_magnbands}
	\end{equation}
	In matrix form, it can be written compactly as
	\begin{equation}
		\mathcal{H}=-t\sum_{\mathbf{k}\in\mathrm{MBZ}}(d^\dag_{\mathbf{k}},\ldots,\;d^\dag_{\mathbf{k}+(n-1)\Phi\hat{1}})\;\mathcal{G}_n
		\begin{pmatrix}
		d_{\mathbf{k}} \\
		\vdots\\
		d_{\mathbf{k}+(n-1)\Phi\hat{1}}
		\end{pmatrix},
		\label{final_momentum_hamiltonian_2dtb}
	\end{equation}
	where
	\begin{equation}
		\mathcal{G}_n=
		\begin{pmatrix}
		2\cos{(k_1)} & e^{-ik_2} & 0 & \dots & e^{ik_2} \\
		e^{ik_2} & 2\cos(k_1+\Phi) & e^{-ik_2} & 0 & \dots \\
		0 & e^{ik_2} & \ddots & \ddots & \ddots\\
		\vdots & 0 & \ddots & \ddots & e^{-ik_2}\\
		e^{-ik_2} & \vdots & \ddots & e^{ik_2} & 2\cos(k_1+(n-1)\Phi)\\
		\end{pmatrix}.
		\label{matrix_n_2dtb}
	\end{equation}	
	This matrix depends on the flux, it has size $n\times n$, and its eigenvalues, for each value of $\mathbf{k}$, provide the energy spectrum of the model. It is immediate to check that the result coincides with the one obtained using directly a gauge, such as the Landau gauge. 
	A simple check can be done in the so-called $\pi-$flux case, where $(m,n)=(1,2)$. Here the matrix is
	\begin{equation}
		\frac{\mathcal{G}_2}{2}
		=
		\begin{pmatrix}
		\cos{k_1} & -\cos{k_2} \\
		-\cos{k_2} & -\cos{k_1}
		\end{pmatrix} 
	\end{equation}
	and the associated spectrum
	\begin{equation}
		E_\mathbf{k}=\pm 2t\sqrt{\cos^2k_1+\cos^2k_2},
		\label{2d_tb_spectrum}
	\end{equation}
	recovering the known $2d$ analytical result \cite{Harper,Affleck-Marston}. For general values of the magnetic fields, 
	i.e. for generic $n$ and $m$, we checked that the spectrum of $\mathcal{G}_n$ is the correct one, e.g. by comparison with the exact diagonalization of Eq. \eqref{hofstadter_hamiltonian}.
	
	\subsubsection{Symmetric construction}
	We present here the rewriting of Eq. \eqref{hofstadter_hamiltonian} using the symmetric construction of Sec. \ref{symm_lattice_construction}. The symmetry considerations leading to the definition of the MBZ still hold, the only difference is that now the size of the lattice has to be $N=2\kappa n$, with $\kappa\in\mathbb{N}$.
	The gauge field for the sites $\mathbf{r}\neq\mathbf{r}_{B,i}$ is now rewritten using Eqs. \eqref{A0_symm_lattice} and \eqref{A1_symm_lattice} , with 
	\begin{equation}
		\bar{F}_{12}=-\Phi(N-r_2),\qquad\bar{F}_{21}=\Phi(N-r_1),\qquad r_i<N.
	\end{equation}
	For the links at the boundary sites $\mathbf{r}_{B,i}$ we use the Eqs. \eqref{symm_PBC_boundaryA0}, \eqref{symm_PBC_boundaryA1} with the gauge invariant loops $\bar{f}_i$ of Eq. \eqref{loops_2dhofst}. As in the asymmetric case, the boundary terms have the same functional form of the bulk ones, because of the assumption on the size $N$. Going into momentum space, the MBZ is 
	\begin{equation}
		\mathrm{MBZ:}\hspace{0.5cm}\bigg[-\frac{\pi}{2n},\frac{\pi}{2n}\bigg)\times\bigg[-\frac{\pi}{2n},\frac{\pi}{2n}\bigg).
		\label{2d_symm_MBZ}
	\end{equation}
	Introducing the gauge invariant operators $d^\dag_{\mathbf{r}},\;d_{\mathbf{r}}$ as in Eq. \eqref{gauge_inv_fermions} and the reduced magnetic field $\tilde{\Phi}\equiv \Phi/2$, the Hamiltonian can be rewritten as
	\begin{equation}
		\mathcal{H}=-t\sum_{\mathbf{k}\in\text{MBZ}}\sum_{\lambda,\tau}[e^{-i(k_1+\tau\tilde{\Phi})}d^\dag_{\mathbf{k}+\tau\tilde{\Phi}\hat{1}+(\lambda+1)\tilde{\Phi}\hat{2}}d_{\mathbf{k}+\tau\tilde{\Phi}\hat{1}+\lambda\tilde{\Phi}\hat{2}}+e^{-i(k_2+\lambda\tilde{\Phi})}d^\dag_{\mathbf{k}+\tau\tilde{\Phi}\hat{1}+\lambda\tilde{\Phi}\hat{2}}d_{\mathbf{k}+(\tau+1)\tilde{\Phi}\hat{1}+\lambda\tilde{\Phi}\hat{2}}+\text{h.c.}].
		\label{symmconstr_hamiltonian_momentumspace}
	\end{equation} 
	using the two magnetic band indices $\tau,\lambda=\{0,\ldots,2n-1\}$. The associated matrix turns out to be of size $(2n)^2\times(2n)^2$, which has to be compared with the $\mathcal{G}_n$, of size $n\times n$, obtained with the asymmetric construction. Being the energy spectrum the same, the main difference is in the definition of the MBZ, as we are going to discuss at the end of the Section.
	
	\subsection{\label{3d}The $3d$ model}
	The analysis done in $2d$ can be extended to the $3d$ model. We consider a cubic lattice of size $V=N^3$, with an isotropic magnetic field $\mathbf{B}=\Phi(1,1,1)$. 
	Different orientations of the magnetic field, such as $\mathbf{B}=\Phi(1,0,0)$, produce different results, but the method is the same and for convenience of exposition we limit ourselves to the isotropic case. In the following we will show how to retrieve the spectrum of the model within our formalism.
	
	\subsubsection{Asymmetric construction}
	The non-trivial components of the field strength tensor are $F_{21}=F_{32}=\Phi$ and $F_{31}=-\Phi$. We use Eq. \eqref{compact_gaugefield_rewriting} to express the vector potential $\mathbf{A}(\mathbf{x})$. The relevant strip variables are
	\begin{equation}
	\bar{F}_{21}=-\bar{F}_{31}=\Phi(N-r_1),\qquad\bar{F}_{32}=\Phi(N-r_2),\qquad r_i<N.
	\end{equation}
	The functional form of the loops $\bar{f}_i$ can be determined by imposing the constraints on the proper flux per plaquette at the boundary sites, as for the $2d$ case. By using the definition in Eq. \eqref{boundarylinks_higherdim}, we obtain the loops
	\begin{equation}
		\bar{f}_1(r_2,r_3)=\Phi N(r_2-r_3)+\vartheta_1,\qquad\bar{f}_2(r_1,r_3)=N\Phi(r_3-r_1)+\vartheta_2,\qquad\bar{f}_3(r_1,r_2)=N\Phi(r_1-r_2)+\vartheta_3.
		\label{loops_3dhofst}
	\end{equation}
	As before, we consider the case $\vartheta_1=\vartheta_2=\vartheta_3=0$.
	Introducing directly the operators in Eq. \eqref{gauge_inv_fermions} and the MBZ
	\begin{equation}
	\text{MBZ}:\;\bigg[-\frac{\pi}{n},\frac{\pi}{n}\bigg)\times\bigg[-\frac{\pi}{n},\frac{\pi}{n}\bigg)\times\bigg[-\pi,\pi\bigg),
	\label{3d_mbz_iso}
	\end{equation}
	we split the structure of the Hamiltonian in magnetic bands, labeled by two indices $\lambda,\;\tau\in\{0,1,\ldots,n-1\}$:
	\begin{align}
		\mathcal{H}=-t\sum_{\mathbf{k}\in\text{MBZ}}\sum_{\lambda,\tau}&\bigg[2\cos(k_1+\lambda\Phi)d^\dag_{\mathbf{k}+\lambda\Phi\hat{1}+\tau\Phi\hat{2}}d_{\mathbf{k}+\lambda\Phi\hat{1}+\tau\Phi\hat{2}}+e^{-i(k_2+\tau\Phi)}d^\dag_{\mathbf{k}+\lambda\Phi\hat{1}+\tau\Phi\hat{2}}d_{\mathbf{k}+(\lambda+1)\Phi\hat{1}+\tau\Phi\hat{2}}\\
		&+e^{-ik_3}d^\dag_{\mathbf{k}+\lambda\Phi\hat{1}+\tau\Phi\hat{2}}d_{\mathbf{k}+(\lambda-1)\Phi\hat{1}+(\tau+1)\Phi\hat{2}}+\text{h.c.}\bigg]\nonumber.
	\end{align}
	The associated matrix has size $n^2\times n^2$. We verified that the spectrum of this Hamiltonian coincides with the known one in literature \cite{Burrello-Trombettoni1}. A simple analytical check can be done in the $\pi-$flux case, where $(m,n)=(1,2)$. Here the matrix is (factorizing an overall factor of 2)
	\begin{equation}
		\mathcal{G}_2=
		\begin{pmatrix}
		\cos{k_1} & 0 & \cos{k_2} & \cos{k_3} \\
		0 & \cos{k_1} & \cos{k_3} & -\cos{k_2} \\
		\cos{k_2} & \cos{k_3} & -\cos{k_1} & 0 \\
		\cos{k_3} & -\cos{k_2} & 0 & -\cos{k_1}
		\end{pmatrix}=\cos{k_1}\;\sigma_z\otimes\mathbb{1}_2+\cos{k_2}\;\sigma_x\otimes\sigma_z+\cos{k_3}\;\sigma_x\otimes\sigma_x
	\end{equation}
	whose eigenvalues are
	\begin{equation}
		\lambda_{1,2}(\mathbf{k})=\pm\sqrt{\cos^2{k_1}+\cos^2{k_2}+\cos^2{k_3}}.
	\end{equation}
	The full spectrum is related to them via
	\begin{equation}
		\frac{E(\mathbf{k})}{2t}=\lambda_{1,2}(\mathbf{k}),
	\end{equation}
	reproducing exactly the dispersion relation in \cite{Burrello-Trombettoni1}. 

	\subsubsection{Symmetric construction}
	One can proceed as in the $2d$ case, the only computational difference being represented by the size of the lattice, which now has to be $N=3\kappa n$, with $\kappa\in\mathbb{N}$. 
	The gauge field components are
	\begin{equation}
		A_1=\phi(\mathbf{r}+\hat{1})-\phi(\mathbf{r})+\frac{2\bar{F}_{13}+\bar{F}_{12}}{3},
	\end{equation}
	\begin{equation}
		A_2=\phi(\mathbf{r}+\hat{2})-\phi(\mathbf{r})+\frac{2\bar{F}_{21}+\bar{F}_{23}}{3},
	\end{equation}
	\begin{equation}
		A_3=\phi(\mathbf{r}+\hat{3})-\phi(\mathbf{r})+\frac{2\bar{F}_{32}+\bar{F}_{31}}{3},
	\end{equation} 
	with the boundary links that can be immediately obtained through the proper symmetrization of Eq. \eqref{boundarylinks_higherdim}.
	Going into momentum space, the resulting MBZ is 
	\begin{equation}
		\text{MBZ}:\hspace{0.3cm}\bigg[-\frac{\pi}{3n},\frac{\pi}{3n}\bigg)\times\bigg[-\frac{\pi}{3n},\frac{\pi}{3n}\bigg)\times\bigg[-\frac{\pi}{3n},\frac{\pi}{3n}\bigg).
		\label{symm_MBZ}
	\end{equation}
	Introducing the gauge invariant operators $d^\dag_{\mathbf{r}},\;d_{\mathbf{r}}$ as in Eq. \eqref{gauge_inv_fermions} and the reduced magnetic field $\tilde{\Phi}\equiv \Phi/3$, the Hamiltonian can be rewritten as
	\begin{align}
		\nonumber
		\mathcal{H}=-t\sum_{\tau,\epsilon,\lambda}\sum_{\mathbf{k}\in\text{MBZ}}\bigg[&d^\dag_{\mathbf{k}+\tau\tilde{\Phi}\hat{1}+(\epsilon+1)\tilde{\Phi}\hat{2}+(\lambda-2)\tilde{\Phi}\hat{3}}d_{\mathbf{k}+\tau\tilde{\Phi}\hat{1}+\epsilon\tilde{\Phi}\hat{2}+\lambda\tilde{\Phi}\hat{3}}e^{-i(k_1+\tau\tilde{\Phi})}\\
		\nonumber
		&+d^\dag_{\mathbf{k}+(\tau-2)\tilde{\Phi}\hat{1}+\epsilon\tilde{\Phi}\hat{2}+(\lambda+1)\tilde{\Phi}\hat{3}}d_{\mathbf{k}+\tau\tilde{\Phi}\hat{1}+\epsilon\tilde{\Phi}\hat{2}+\lambda\tilde{\Phi}\hat{3}}e^{-i(k_2+\epsilon\tilde{\Phi})}\\
		&+d^\dag_{\mathbf{k}+(\tau+1)\tilde{\Phi}\hat{1}+(\epsilon-2)\tilde{\Phi}\hat{2}+\lambda\tilde{\Phi}\hat{3}}d_{\mathbf{k}+\tau\tilde{\Phi}\hat{1}+\epsilon\tilde{\Phi}\hat{2}+\lambda\tilde{\Phi}\hat{3}}e^{-i(k_3+\lambda\tilde{\Phi})}+\text{h.c.}\bigg],
	\end{align}
	with the help of three magnetic band indices $\tau,\epsilon,\lambda=\{0,\ldots,3n-1\}$. The size of the associated matrix is $(3n)^3\times(3n)^3$, much larger than the one obtained with the asymmetric construction. Obviously, the spectra associated to the same pair $(m,n)$ are found to coincide.

	We are now ready to compare the two constructions applied to the Hofstadter model. First, we remind that using the Hasegawa gauge \cite{Hasegawa} in $3d$ (or the Landau gauge \cite{Landau} in $2d$) one has to diagonalize, for each $\mathbf{k}$ belonging to the MBZ, a matrix $n \times n$. If one uses a different gauge, the matrix to be diagonalized may be of larger size, and, at the same time, the MBZ also changes.
	What does not change is the energy spectrum. 
	Let us now discuss the results obtained using our formalism in which a choice of the gauge is not done. In $2d$, with the asymmetric construction we obtained a matrix in momentum space of size $n\times n$, where the MBZ is Eq. \eqref{2d_asymm_MBZ}. With the symmetric construction we symmetrized the MBZ, as showed in Eq. \eqref{2d_symm_MBZ}, and the band structure of the Hamiltonian, but the price to pay is in the dimensionality of the matrix, of size $(2n)^2\times(2n)^2$, larger than the asymmetric one. The same considerations hold for the $3d$ case, with different dimensions of the matrix, respectively $n^2\times n^2$ and $(3n)^3\times(3n)^3$, and the definitions of MBZ, given by Eqs. \eqref{3d_mbz_iso} and \eqref{symm_MBZ}. One concludes that the MBZ depends, in our formalism, on the chosen construction. Furthermore the most symmetric MBZ, in which the $x,y,z$ axis enter equally -- as one would expect since the magnetic field is isotropic -- is given by the symmetric contruction
	\footnote{We remark that from the textbook definition of MBZ one can see that its volume is unique for each dimensionality \cite{Lifschitz-Pitaevski}. Focusing on the $2d$ case, this is because, denoting with $\hat{T}_{\hat{j}}$ the generators of the magnetic translational group, the minimal integer doublet $(a,b)$ such that $[\hat{T}_{a\hat{0}},\hat{T}_{b\hat{1}}]=0$ defines the magnetic unit cell in real space. For the asymmetric construction we have $(a,b)=(2,1)$, while in the symmetric one $(a,b)=(4,4)$, therefore the minimal one is the first one.}. Therefore, if the symmetry in momentum space has to be preserved, it is more convenient to use the symmetric construction.	If, at variance, one wants to reduce the dimension of the matrix to be diagonalized, e.g. for numerical purposes, then the asymmetric construction is more suitable.
	
	\section{\label{discussion_applications}{Applications of the formalism}}
	
	In the previous Sections, after having introduced the main ideas of the reformulation, we applied it to different cases: a single particle in a static magnetic field, the pure lattice gauge theory (without dynamical matter), the Hofstadter model (where the magnetic field is static) and QED (where there are both dynamical matter and gauge fields). The reformulation can be applied as well to other models, such as the Schwinger-Thirring model, where the matter is interacting with a term $\mathcal{L}_{int}\propto(\bar{\psi}\gamma^\mu\psi)^2$, the Gross-Neveau model or bosonic QED \cite{frishman}. For example, in the case of the Thirring interaction, the Lagrangian would be
	\begin{equation}
		\mathcal{L}[\bar{F}_{\mu\nu},\psi',\bar{\psi'}]=\mathcal{L}_{\text{QED}}[\bar{F}_{\mu\nu},\psi',\bar{\psi'}]+g(\bar{\psi}'\gamma^\mu\psi')^2,
	\end{equation}
	where $\mathcal{L}_{\text{QED}}[\bar{F}_{\mu\nu},\psi',\bar{\psi'}]$ is the reformulated Lagrangian in Eq. \eqref{QED_lagrangian_newvars}. Moreover, the reformulation could be applied to models in ladder geometries, which are nowadays subject of considerable attention \cite{Science_Barbiero_2019,Tagliacozzo_PRL_2020,Santos_PRL_2020,Borla_PRL_2020}. The most interesting theories are the ones in which the matter is interacting in presence of a magnetic field, or, even more, lattice and continuum gauge theories where the gauge fields are dynamical. 
	In the following paragraphs, we discuss how the present formalism may be advantageous in both cases.
	
	For fermions in a static magnetic field, interactions give rise to the so-called Hubbard-Hofstadter model, which is considerably more difficult to analyze.  
	Our reformulation provides an alternative path, arguably more suitable as it preserves gauge invariance exactly, to study these models under certain approximation schemes, like mean-field.
	Indeed, the order parameters that one may introduce in the (non-magnetic) Hubbard model are clearly not gauge invariant. At variance, using the $d$ operators one can construct gauge invariant order parameters, whose self-consistency has to be checked. In a similar way the correlation functions of the Hubbard model, when an approximation is used, are expected to be gauge dependent. However, if one determines, even in an approximate way, correlation functions of the $d$ operators, they will be gauge invariant. We believe this constitues a promising line of research.
	
	The situation is even more relevant for standard gauge theories, which have dynamical gauge fields.
	The application of a naive mean-field approximation leads to a self-consistent equation for a non gauge-invariant quantity, 
	which is in tension with Elitzur's theorem \cite{Wilson,Elitzur}. On the lattice, subsequent efforts were able to fix these drawbacks by introducing a generalized mean-field procedure, where several ``mean-fields" for each gauge degree of freedom are introduced \cite{DROUFFE}. In general, this procedure appears rather cumbersome to be implemented and not easily extendable to the continuum. 
	The present reformulation provides an alternative starting point for a mean-field approximation where a self-consistent equation for the targeted order parameter can be written in agreement with Elitzur's theorem. The main challenge consists in identifying suitable order parameters in terms of $\bar{F}_{\mu\nu},\;\psi'$. 
	More work on these lines is actively being pursued.

	\section{\label{conclusions}Conclusions}
	In this paper we gave a reformulation of abelian gauge theories in terms of gauge invariant fields 
	(GIF). In particular, we discussed how to split the gauge field $A_\mu$ into its gauge invariant part, represented by $F_{\mu\nu}$, and its gauge covariant one, enclosed in the vertex variables $\phi$. From the field stress tensor, we have introduced the plaquette strips $\bar{F}_{\mu\nu}$
	in order to define a set of independent field variables $\{\bar{F}_{\mu\nu},\phi\}$, whose determination is the main goal of our reformulation. For periodic boundary conditions, these variables are suplemented by loops $\bar{f}_\mu$ that wrap around the different directions of space-time.

	The construction was first developed on the lattice and in $(1+1)$ dimensions, which provides the basic building block for the generalization to arbitrary dimensions and to the continuum limit. The choice of how to make these constructions is not unique
	and here we explored two of them, that we called asymmetric and symmetric constructions. All possible constructions are related by gauge transformations and, therefore, are physically equivalent. We stress, however, that a gauge is not fixed and we deal only with fields that are independent of any possible chosen gauge. The procedure is performed at finite volume and we have used open (OBC) and periodic (PBC) boundary conditions, which fit well within the formalism.

	This kind of constructions arises naturally, as a change of variables, in the Lagrangian formalism. However, performing their canonical quantization, it is predicted to be an arduous task due to the presence of non-local kinetic terms. Despite that, we emphasize that the non-locality of the Lagrangian formalism does not break unitarity nor Lorentz invariance. Furthermore, we showed through two examples, i.e. charged particle in an external magnetic field and in the Hofstadter model, that the same kind of construction can be applied in the Hamiltonian formalism.
	
	From the example provided by the Hofstadter model it becomes clear that different choices on the constructions have practical implications. The Hamiltonian diagonalization can be reduced to the diagonalization of matrices of finite size for every value of the momentum. The asymmetric construction leads to smaller matrices but an asymmetric magnetic Brillouin zone (MBZ). At variance, the symmetric constructions implies the diagonalization of larger matrices but produces a symmetric MBZ.
	
	In the literature, there are similar efforts of describing gauge theories solely in terms of gauge invariant fields. The method presented in  \cite{gaugeinv_rudolph_1,gaugeinv_rudolph_2,gaugeinv_rudolph_3,gaugeinv_rudolph_4} recombines properly the matter and gauge fields in order to rewrite QED in terms of gauge invariants, and quantize the theory in terms of them. Difficulties arise due to the presence of non-local quantities in the quantization procedure \cite{gaugeinv_rudolph_2,gaugeinv_rudolph_4}. In this aspect this is analogous to our rewriting, as the plaquette strips interact non-locally. Moreover, the authors underline that, within their approach, the fermions and the photons are no longer fundamental fields \cite{gaugeinv_rudolph_2}. The new degrees of freedom are the currents of the theory, regarding the matter, and a couple of covector and complex scalar fields, regarding the gauge part. In our rewriting the new degrees of freedom are different: we combined the vertex variables with the matter fields to obtain	degrees of freedom that remain fermionic but are also now gauge invariant. Some aspects of our reformulation are also shared with  \cite{Kaplan-Stryker,Haase-Dellantonio,Bender-Zohar} where the plaquettes terms, on the lattice, are used to replace the links associated to the gauge field $A_\mu$. The main difference lies, again, in the definition of the matter variables, in Eqs. \eqref{gauge_inv_newfields} and \eqref{gauge_inv_fermions} as just described. In our procedure, this leads to non-local interactions between gauge and matter fields. The removal of non-locality, of a similar form, was solved in \cite{Bender-Zohar}, thanks to the introduction of new variables. A future interesting step would be to understand if it is possible to introduce further variables, in a similar way, in order to make the matter-gauge interacting term of our theory local as well.

	The reformulation presented here has been applied to gauge theories with an abelian symmetry group. We expect that the generalization to non-abelian gauge theories is possible by following the same lines presented here, and we do not anticipate specific problems related to the non-abelian nature of the symmetry group. At variance, we think that the generalization to LGT on non-bipartite lattices would not be straightforward. Both issues will be subject of future studies. Moreover, in view of possible developments, we observe that this formalism could be useful to investigate phase diagrams of gauge theories within a mean-field framework, both in the continuum and on the lattice.
	
	Finally, we would like to stress that both classical and quantum simulations of gauge theories may considerably profit from the reformulation presented 
	here. Clearly, reducing the number of degrees of freedom is potentially interesting in both cases. For the case of quantum simulators there is no longer a local symmetry to be implemented, as it was used to decouple the non-physical fields. From the point of view of the reduction of the number of degrees of freedom, this formalism is on a similar footing with maximal gauge fixings \cite{Creutz1977}. The difference lies in the fact that our reformulation allows for the identification of a new matter field $\psi'$, expressed in terms of the variables $\phi$ which are integrated out, and new fields $\bar{F}_{\mu\nu}$. Both $\psi'$ and $\bar{F}_{\mu\nu}$ are gauge invariant and the reformulated theory is expressed in terms of them: ${\cal L}={\cal L}(\psi',\bar{F}_{\mu\nu})$. In this, which is for all practical purposes a trading of difficulties, the final Lagrangian has non-local terms and the construction of the Hamiltonian could be a non-trivial step. Depending on the form of the final theory, this may provide the starting point of approximate methods, in which correlation functions and order parameters are gauge invariant by construction. Ultimately, it will be the success of performing sensible approximations of interacting lattice field theories that will show whether the reformulation presented here is useful. More work on this topic is currently actively pursued.
	\\\\
	\indent\textit{Acknowledgements:} We thank M. Burrello, M. Dalmonte, A. Galvani, G. Giachetti, L. Lepori and U.-J. 
	Wiese for very useful discussions.

	\bibliography{biblio}% Produces the bibliography via BibTeX.

\end{document}